\documentclass[journal]{IEEEtran}
\usepackage{ifpdf,cite,epsf,psfrag,epsfig}
\usepackage[cmex10]{amsmath}
\usepackage{mathrsfs, amsthm,amsfonts, latexsym, amssymb,algorithmic,algorithm}
\usepackage[mathscr]{eucal}
\usepackage[caption=false,font=scriptsize]{subfig}
\usepackage{multicol,multirow,color}
\usepackage[breaklinks]{hyperref}
\usepackage[hyphenbreaks]{breakurl}

\begin{document}
\title{Data-Driven Load Modeling and Forecasting of Residential Appliances}
\author{Yuting Ji, Elizabeth Buechler, and Ram Rajagopal
\thanks{This work was supported in part by the NSF under award ECCS-1554178, in part by the DOE SunShot Office under award DE-EE00031003, in part by the DOE ARPA-E under award DE-AR0000697, and in part by a Stanford Graduate Fellowship.}
\thanks{Y. Ji and R. Rajagopal are with the Department of Civil and Environmental Engineering, Stanford University, CA, 94305, USA (e-mail: yutingji@stanford.edu; ramr@stanford.edu).}
\thanks{E. Buechler is with the Department of Mechanical Engineering, Stanford University, CA, 94305, USA (email: ebuech@stanford.edu).}
}

\maketitle
\begin{abstract}
The expansion of residential demand response programs and increased deployment of controllable loads will require accurate appliance-level load modeling and forecasting. This paper proposes a conditional hidden semi-Markov model to describe the probabilistic nature of residential appliance demand, and an algorithm for short-term load forecasting. Model parameters are estimated directly from power consumption data using scalable statistical learning methods. Case studies performed using sub-metered 1-minute power consumption data from several types of appliances demonstrate the effectiveness of the model for load forecasting and anomaly detection.

\end{abstract}


\IEEEpeerreviewmaketitle

\section{Introduction}\label{sec:intro}

The deployment of smart grid technologies, such as advanced metering, smart appliances, and automated load control systems in residential applications will change the way demand side management (DSM) is performed. These technologies will give utilities and consumers better visibility into and control of energy demand, allow customers to respond to signals from utilities more quickly, and facilitate the integration of distributed energy resources (DERs). To utilize these capabilities for DSM, utilities need to have a better understanding of how consumers use electricity, particularly at the appliance level. For example, many current demand response (DR) programs \cite{Navigant17} require specific types of appliances, such as air conditioners and water heaters, to modify their consumption behavior in response to DR events. Additionally, increasing DER integration may require utilities to utilize various flexibility options to maintain grid reliability, including load control at the individual appliance or aggregate level. The performance of these emerging DSM applications will rely heavily on the ability to model and forecast individual residential loads at high resolution.


Modeling and forecasting of residential demand is challenging for several reasons. At the meter level, emerging trends such as distributed solar generation, behind-the-meter storage, improved energy efficiency, and electric vehicle (EV) adoption make forecasting more complex. At the appliance level, load profiles are dependent on behavioral and environmental factors and therefore display significant uncertainty, volatility and variability. Appliance characteristics and consumption patterns can also vary significantly from one household to another. A data-driven approach to modeling, utilizing high resolution appliance-level or circuit-level consumption data, is required to address these challenges. Although residential electricity consumption is most commonly recorded only at the meter level, the deployment of sub-metering technologies is expected to increase as low cost solutions become available \cite{DOE17Submeter}.  


The purpose of this study is to develop high resolution models of residential appliances from real power data. In particular, we aim to develop \textit{generic, scalable, and robust} load models to support DSM applications from a utility perspective. 

 \begin{figure}
 \includegraphics[width=.48\textwidth]{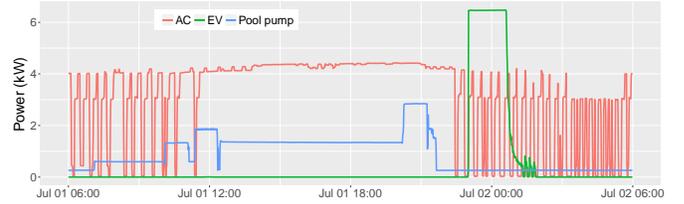}
 \caption{Power trajectories of the top three electricity consuming appliances (Data source: 2017 Pecan Street Home 5357 \cite{PecanSt17Dataport}). }\label{fig:appliances}
 \end{figure}
 
Our model is designed based on two key features observed in high resolution load data: (1) a finite set of discrete operating states and (2) random time durations spent in each state. These two features can be observed in Fig.~\ref{fig:appliances}, which shows real power trajectories for three different types of appliances. Although the three appliances in this example have very different behaviors, each is characterized by step changes in power levels and random time periods between state changes.  

To capture these stochastic dynamics, we propose a data-driven approach based on a conditional hidden semi-Markov model (CHSMM). The model assumes that the physical operating states of each appliance are unobservable, but are reflected in the power consumption. While it is assumed in a hidden Markov model (HMM) that the duration distribution is geometric, this assumption is relaxed in a hidden semi-Markov model (HSMM) such that the duration distribution can be explicitly defined. This allows for more accurate modeling of appliances whose state transitions cannot be described well by a Markov process. Influential factors, such as outdoor temperature and time of day, are integrated into the model as exogenous variables. Estimating parameters of a CHSMM is not trivial, especially as the dimensions of the state and duration spaces and time horizon of the training set become large. Also, computational cost will increase significantly with the complexity of the distribution used to describe the state transition probabilities and observation emission probabilities. To address this scalability issue, we use regression to estimate the transition and emission distributions.

\subsection{Summary of Contributions}
Our paper makes the following contributions. First, we develop a probabilistic load model based on a CHSMM, which is broadly applicable to a variety of residential appliances. Model parameters are estimated directly from data through scalable and robust statistical learning methods. Second, we propose a short-term load forecasting algorithm using the learned model. Finally, the effectiveness of the model for load forecasting of individual appliances and aggregations of appliances and for anomaly detection is demonstrated using real-world data.
\subsection{Related Work}\label{sec:lit}

Load modeling of residential appliances has been extensively studied in the literature for various purposes. Here we highlight some of the most relavent work. For broader interest, see reviews \cite{swan2009modeling,grandjean2012review} and the references therein.

Approaches are generally either physics-based, statistical, or utilize a combination of both methods. Numerous detailed physics-based models derived from first principles have been developed and used for residential end-use simulation and control~\cite{jin2017foresee,crawley2001energyplus,fuller2012modeling}. However, the accuracy of physics-based models may suffer due to nonidealities and heterogeneity in appliance characteristics and consumer behavior. 

Various statistical approaches have been developed for non-intrustive load monitoring \cite{KimEtal11SIAM_HSMM,GuoWangKashani15TPS,KongDongHill16TSG_HHMM}. Several variants of a factorial hidden Markov model are considered in \cite{KimEtal11SIAM_HSMM} for load disaggregation. In particular, a conditional factorial hidden semi-Markov model was shown to produce the best performance. In \cite{GuoWangKashani15TPS}, the authors develop a second order model called a explicit-duration hidden Markov model with differential observations. A hierarchical hidden Markov model was proposed in \cite{KongDongHill16TSG_HHMM} to represent appliances with multiple operating states. Although these studies develop variants of a HSMM for load modeling, the application is mainly for load disaggregation which usually focuses on a single household. Our goal, on the other hand, is to develop a generic model which can be applied to a broad set of appliances from different households with a particular interest in short-term load prediction. 

Other studies have taken a bottom-up approach to model residential load at the household, substation, or distribution network level, using both statistical and physics-based approaches \cite{walker1985residential,capasso1994bottom,widen2010high,muratori2013highly}. These models explicitly model occupancy and different occupant activities using probabilistic methods, but rely on consumer time-of-use surveys to characterize ``typical'' use patterns. Thus, these methods cannot accurately model use patterns at sufficiently high resolution or account for heterogeneity within the population. Instead, our approach is to implicitly learn use patterns through model parameter estimation, using only historical power consumption data.

There was been limited work on load forecasting at the appliance level. The authors of \cite{barbato2011forecasting} present an appliance load forecasting system to predict the device state, time of use, and state duration. However, all predictions are based on model-free empirical statistics. In \cite{candanedo2017data}, multiple linear regression, support vector machines with a radial kernel, random forests, and gradient boosting machines are evaluated for appliance load prediction. 

\section{Appliance Load Characterization}\label{sec:pre}
\begin{figure}\centering
\subfloat[]{\includegraphics[width=.48\textwidth]{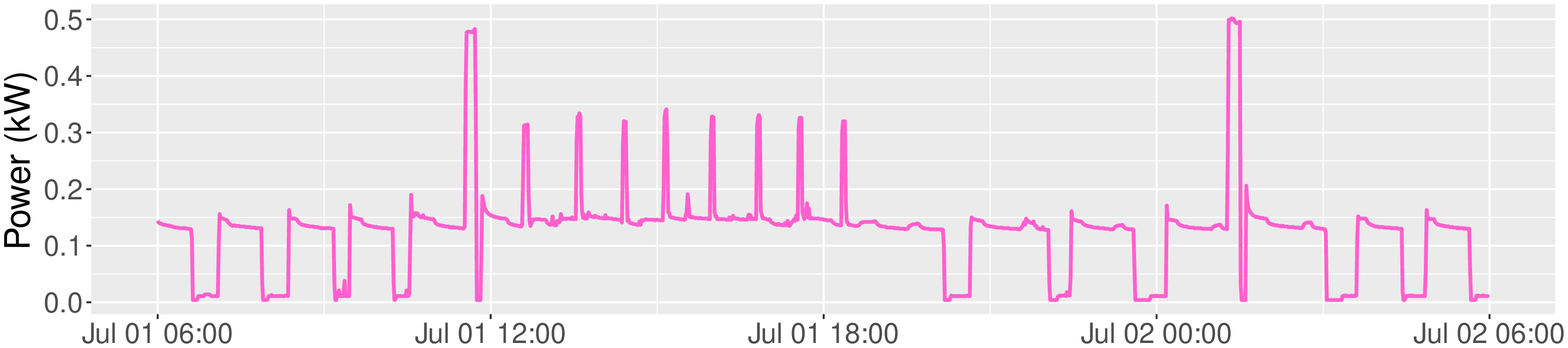}}\\
\subfloat[]{\includegraphics[width=.48\textwidth]{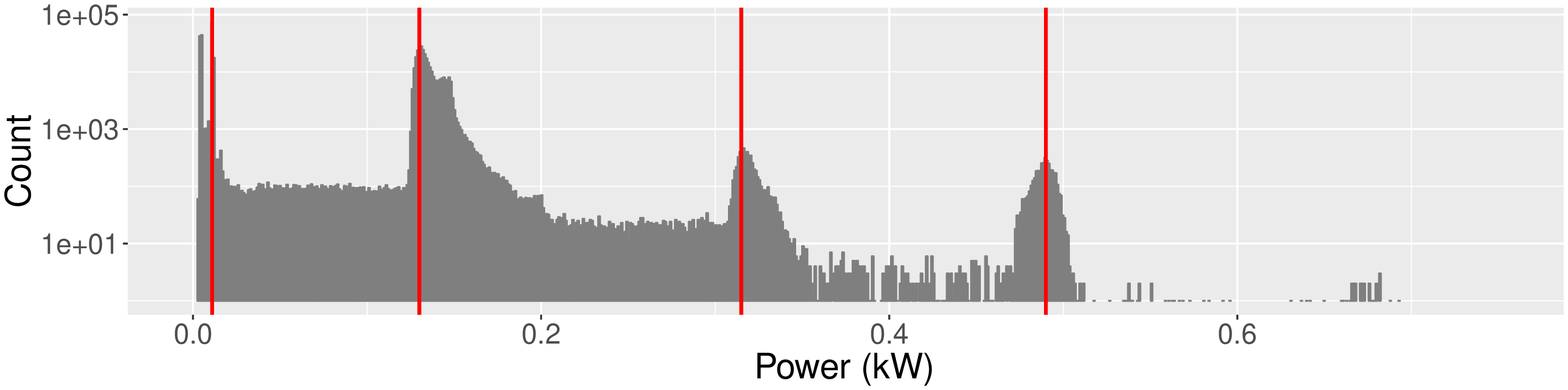}}    
    \caption{Refrigerator from Home 871~\cite{PecanSt17Dataport}: (a) sample real power trajectory over 24-hour period with 1-minute resolution, and (b) real power histogram of the selected refrigerator over year 2017.}
    \label{fig:fridge}
\end{figure}

The proposed model is based on the following observations of appliance consumption: discrete operating states and random state duration. To illustrate these features, we use the real power data of a selected refrigerator from Pecan Street dataset~\cite{PecanSt17Dataport}. A sample real power trajectory of this refrigerator is shown in Fig.~\ref{fig:fridge}a.

\subsection{Discrete Operating States}
As prior work \cite{hart1992nonintrusive} has identified, most residential appliances are characterized by a finite set of discrete operating states, each associated with a different level of power consumption. This is shown in Fig.~\ref{fig:appliances} and Fig.~\ref{fig:fridge}a by the step changes in the 1-minute power consumption trajectories. The discrete states can also be observed in Fig.~\ref{fig:fridge}b, which shows a histogram of the real power measurements of the selected refrigerator over a year, plotted on a log scale. Four peaks are clearly observed at approximately 5 W, 130 W, 300 W and 500 W, as indicated by the vertical lines. These power levels likely correspond with the operating states of the appliance: compressor OFF, compressor ON, ice making, and defrosting. 
 
The states are not directly observable, but can be inferred from observations of the power level. We apply K-means clustering, an unsupervised learning method, to identify the hidden operating states of the appliance from the power measurements. The number of states can be selected manually or through heuristic methods such as the elbow method\cite{tibshirani2001estimating}. 

\subsection{Duration Analysis}
\begin{figure}\centering
\subfloat[]{\includegraphics[scale=.22]{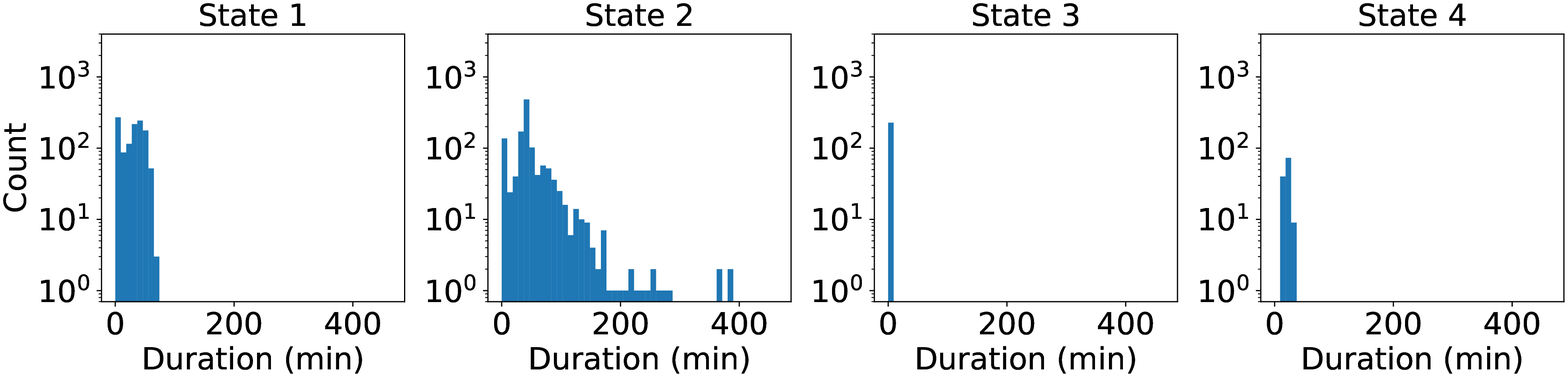}}\\
\subfloat[]{\includegraphics[scale=0.22]{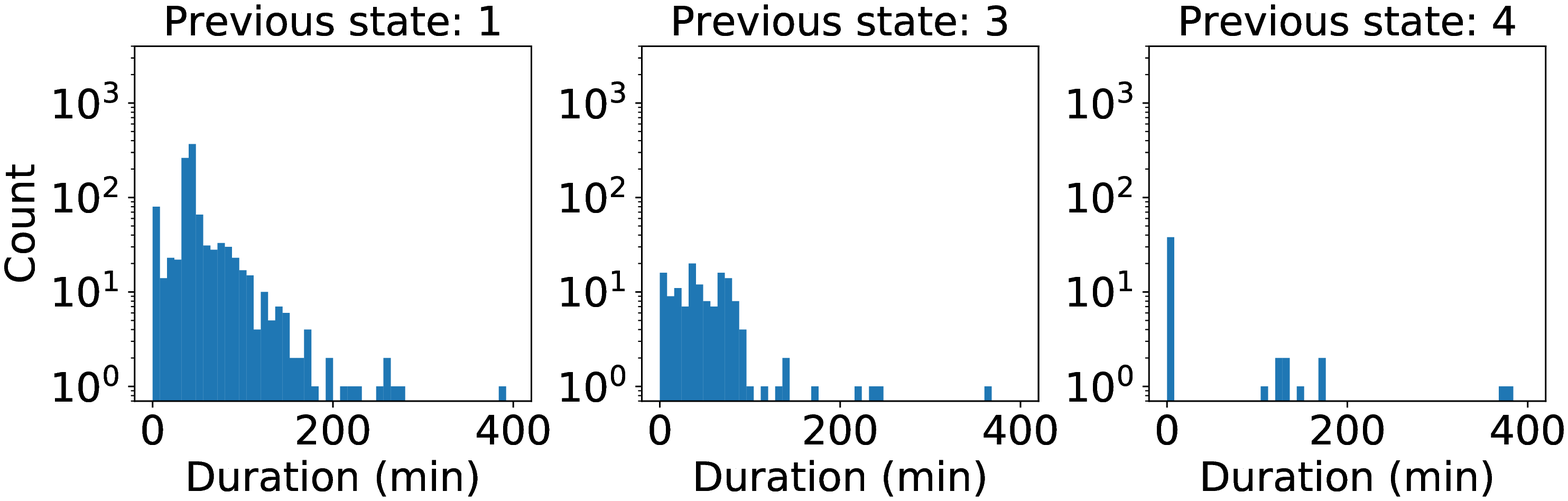}}
\caption{Empirical duration distributions for a four state refrigerator: (a) each histogram shows the marginal duration distribution for a different state, and (b) each histogram shows the duration distribution for state 2, conditioned on a different previous state.}\label{fig:dur_dist}
\end{figure}

The second observation is the randomness of the state duration. First of all, the probability distribution of the duration is state dependent. Histograms of the duration of each identified state of the selected refrigerator are shown in Fig.~\ref{fig:dur_dist}a. The histograms vary significantly by state, such that the duration cannot be accurately described by a single distribution. Second, analysis shows that the duration is also dependent on the previous state. Fig.~\ref{fig:dur_dist}b shows histograms of the duration in state 2, each conditioned on a different previous state. The conditional distributions conditioned on state 3 and 4 are very different from the marginal distribution of state 2 shown in Fig.~\ref{fig:dur_dist}a. Therefore, using the marginal duration distribution alone in a load model may not accurately characterize load behavior.

To account for the complexity of the duration distribution and interdependence of duration and state, we use a HSMM instead of a HMM for load modeling, as the former provides more flexibility in modeling the duration distribution.
\section{Hidden Semi-Markov Model}\label{sec:HSMM}
A HSMM \cite{Yu10AI_HSMM} extends the concept of a HMM \cite{Rabiner89IEEE_HMM} to include the case where each state has a variable duration. The basic idea underlying the HSMM formalism is to augment the generative process of a standard HMM with a random state duration time, drawn from some state-specific distribution when each state is entered. The state remains constant until the duration expires, at which point there is a Markov transition to a new state. This formulation eliminates the implicit geometric duration distribution assumptions in the standard HMM, and thus allows the state to transition in a non-Markovian way. A HSMM is characterized  by the following components: 

(1) The state space $\mathscr{S}=\{S_1,S_2,\cdots, S_{N_S}\}$. We introduce the concept of an epoch to index the transition of states, denoting the state at time $t$ by $x_t$ and at epoch $k$ by $\tilde{x}_k$. In contrast with the time-indexed state $x_t$, self-transition is not allowed for the epoch-indexed state $\tilde{x}_k$, \textit{i.e.}, $\tilde{x}_{k}\neq \tilde{x}_{k+1}$ for all $k$. Let $t^s_k$ and $t^e_k$ denote the starting and ending times of the $k$-th epoch respectively, then $x_t=\tilde{x}_k$, for all  $t\in[t^s_k,t^e_k]$.

(2) The duration space $\mathscr{D}=\{D_1, D_2,\cdots, D_{N_k}\}$, where the number of possible durations $N_k$ is finite. The duration is an integer equal to the number of time intervals occupied by each state. We denote the duration at epoch $k$ by $\tilde{d}_k$, whose value is equal to $t^e_k-t^s_k+1$.

(3) The observation space $\mathscr{O}=\{O_1,O_2,\cdots, O_{N_O}\}$. We denote the observation at time $t$ by $y_t$ and at epoch $k$ by $\tilde{y}_k$. Note that $\tilde{y}_k$ is a vector of length $\tilde{d}_k$, \textit{i.e.}, $\tilde{y}_k =\left({y}_{t_k^s}, {y}_{t_k^s+1}, \cdots, {y}_{t_k^e}\right)$.

(4) The generalized state transition probability matrix $A$, where the generalized state is the double $(\tilde{x}_k,\tilde{d}_k)\in\mathscr{S} \times \mathscr{D}$, and the element in row $(i,l)$, column $(j,m)$ of $A$ is given by
\begin{equation}\label{eqn:proba} \begin{split}
a_{(i,l), (j,m)}=\mathbb{P}[\tilde{x}_{k+1}=S_j, \tilde{d}_{k+1}=D_m|\tilde{x}_{k}=S_i, \tilde{d}_k=D_l],\\
 \forall S_i, S_j\in\mathscr{S}, \forall D_l, D_m \in \mathscr{D}.
\end{split}
\end{equation}

(5) The emission probability distribution $B$, whose probability mass (density) function $b_{i}(y)$ is defined as follows. For the case of finite $N_O$,
\begin{equation}
b(y|i)=\mathbb{P}[{y}_t=y|{x}_t=S_i], \forall S_i\in\mathscr{S}, \forall y\in\mathscr{O}
\end{equation}
For infinite $N_O$,
\begin{equation}\begin{split}
\int_{O_j}^{O_l} b(y|i) dy=\mathbb{P}[O_j \leq {y}_t \leq O_k|{x}_t=S_i], \\ \forall S_i\in\mathscr{S}, \forall O_j, O_l \in \mathscr{O}.
\end{split}
\end{equation}
Note that the emission probability is assumed to be time-independent.

(6) The initial  distribution $\pi$ whose $(i,j)$-th element represents the probability of the initial state being $S_i$ and its duration being $D_j$, \textit{i.e.}
\begin{equation}
\pi_{(i,j)}=\mathbb{P}[\tilde{x}_1=S_i,\tilde{d}_1=D_j], \forall S_i\in\mathscr{S}, \forall D_j\in\mathscr{D}.
\end{equation}

\begin{figure}\centering
\begin{psfrags}
\psfrag{z}[c]{\small $\cdots$}
\psfrag{o1}[c]{\small ${y}_1$}
\psfrag{o2}[c]{\small ${y}_2$}
\psfrag{o3}[c]{\small ${y}_3$}
\psfrag{o4}[c]{\small ${y}_4$}
\psfrag{o5}[c]{\small ${y}_5$}
\psfrag{o6}[c]{\small ${y}_6$}
\psfrag{o7}[c]{\small ${y}_7$}
\psfrag{o8}[c]{\small ${y}_8$}
\psfrag{o9}[c]{\small ${y}_9$}
\psfrag{o10}[c]{\small ${y}_{T}$}
\psfrag{o0}[c]{\small }
\psfrag{s1}[c]{\small $\tilde{x}_1$}
\psfrag{s2}[c]{\small $\tilde{x}_2$}
\psfrag{s3}[c]{\small $\tilde{x}_K$}
\psfrag{d1}[c]{\small $\tilde{d}_1$}
\psfrag{d2}[c]{\small $\tilde{d}_2$}
\psfrag{d3}[c]{\small $\tilde{d}_K$}
\psfrag{t1}[c]{\scriptsize $1$}
\psfrag{t2}[c]{\scriptsize $2$}
\psfrag{t3}[c]{\scriptsize $3$}
\psfrag{t4}[c]{\scriptsize $4$}
\psfrag{t5}[c]{\scriptsize $5$}
\psfrag{t6}[c]{\scriptsize $6$}
\psfrag{t7}[c]{\scriptsize $7$}
\psfrag{t8}[c]{\scriptsize $8$}
\psfrag{t9}[c]{\scriptsize $9$}
\psfrag{t10}[c]{\scriptsize $T$}
\psfrag{t0}[c]{\scriptsize}
\includegraphics[scale=0.8]{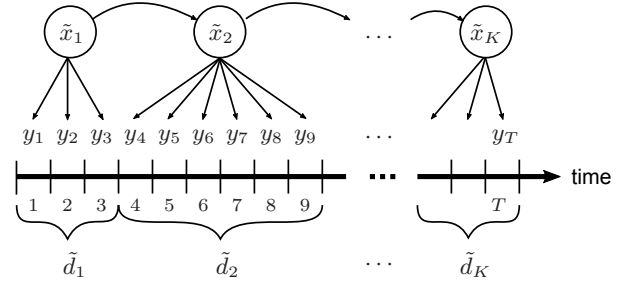}
\end{psfrags}
\caption{Graphical representation of a standard HSMM.}\label{fig:structure}
\end{figure}

An example of the HSMM model described above is shown in Fig.~\ref{fig:structure}. In the $T$-period time horizon, there are a total of $K$ epochs. The first state $\tilde{x}_1$ and its duration $\tilde{d}_1=3$ are selected according to the initial distribution $\pi_{(\tilde{x}_1,\tilde{d}_1)}$. The generalized state $(\tilde{x}_1,\tilde{d}_1)$ produces three observations $\tilde{y}_1=({y}_1, {y}_2,{y}_3)$ following the emission probability $b_{\tilde{x}_1}(\cdot)$. According to the transition probability $a_{(\tilde{x}_1,\tilde{d}_1), (\tilde{x}_2,\tilde{d}_2)}$, the state transits from $\tilde{x}_1$  to $\tilde{x}_2$ which is occupied for $\tilde{d}_2=6$ intervals. In the second epoch, six observations $\tilde{y}_2=({y}_4,{y}_5,\cdots,{y}_9)$ are generated based on $b_{\tilde{x}_2}(\cdot)$. Such transitions continue until the last observation ${y}_T$ is produced by the last state $\tilde{x}_K$ in epoch $K$. Note that this final state $\tilde{x}_K$ may last longer than $\tilde{d}_K$, but we impose a finite horizon T for this example.

\section{Appliance Load Model}
In this section, we first present a general probabilistic model of appliance power consumption. Then a scalable and robust statistical approach is described for model parameter estimation. Finally, we discuss two model variations that improve model accuracy for specific loads.  

\subsection{Conditional HSMM}
As the power consumption of most appliances is highly dependent on external variables such as outdoor temperature and time of day, we propose a CHSMM, where the state transition probabilities and emission probabilities are not constant but conditioned on exogenous variables.

\begin{figure}\centering \begin{psfrags}
\psfrag{x1}[c]{\small $\tilde{x}_{1}$}
\psfrag{d1}[c]{\small $\tilde{d}_{1}$}
\psfrag{x2}[c]{\small $\tilde{x}_{2}$}
\psfrag{d2}[c]{\small $\tilde{d}_{2}$}
\psfrag{x3}[c]{\small $\tilde{x}_{K}$}
\psfrag{d3}[c]{\small $\tilde{d}_{K}$}
\psfrag{y1}[c]{\small ${y}_{t^s_1}$}
\psfrag{y2}[c]{\small ${y}_{t^e_1}$}
\psfrag{y3}[c]{\small ${y}_{t^s_2}$}
\psfrag{y4}[c]{\small ${y}_{t^e_2}$}
\psfrag{y5}[c]{\small ${y}_{t^s_K}$}
\psfrag{y6}[c]{\small ${y}_{t^e_K}$}
\psfrag{z1}[l]{\small $\tilde{z}_{1}$}
\psfrag{z2}[l]{\small $\tilde{z}_{2}$}
\psfrag{z3}[l]{\small $\tilde{z}_{K}$}
\psfrag{w1}[l]{\small $w_{t^s_1}$}
\psfrag{w2}[l]{\small $w_{t^e_1}$}
\psfrag{w3}[l]{\small $w_{t^s_2}$}
\psfrag{w4}[l]{\small $w_{t^e_2}$}
\psfrag{w5}[l]{\small $w_{t^s_K}$}
\psfrag{w6}[l]{\small $w_{t^e_K}$}
\psfrag{dots}[l]{\small $\cdots$}
\includegraphics[width=.48\textwidth]{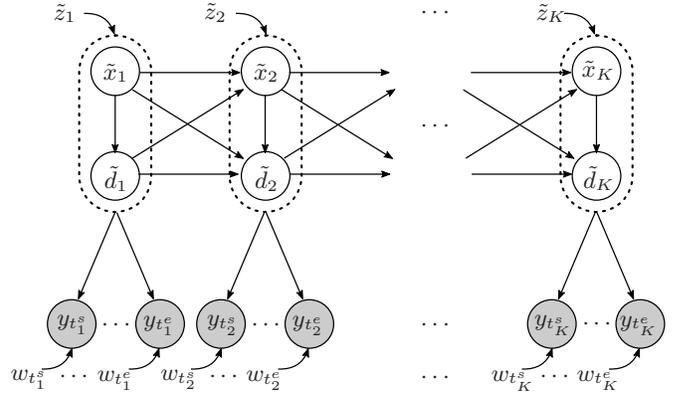}
\end{psfrags}
\caption{Graphical diagram of the proposed appliance load model. Circles represent states and durations, the dashed ovals the generalized states, and the shaded circles the observations. Arrows represent dependencies.}\label{fig:chsmm}\end{figure}

A graphical representation of the proposed CHSMM for appliance load modeling is given in Fig.~\ref{fig:chsmm}, which shows a time segment with $K$ epoches corresponding with $T=t^e_K$ time periods. The notation for states, durations, and observations are the same as for the standard HSMM given in Section~\ref{sec:HSMM}. The CHSMM includes two sets of exogenous variables which are not present in the HSMM. The first set governs the state (and duration) transition between consecutive epoches --- the transition from generalized state $(\tilde{x}_{k-1},\tilde{d}_{k-1})$ at epoch $k-1$ to $(\tilde{x}_k,\tilde{d}_k)$ at epoch $k$ depends on the exogenous variable $\tilde{z}_k$. The second set affects the observation --- the distribution of the observation $y_t$ at time $t$ is conditioned on the current state $x_t$ and exogenous variable $w_t$ at time $t$. It should be noted that both $\tilde{z}_k$ and $w_t$ can be scalars or vectors depending on the number of features. 

We note that the proposed CHSMM is a general form of a HSMM, where independence of state or duration is not assumed. This is different from simplified models such as the \textit{explicit-duration HMM}~\cite{GuoWangKashani15TPS} in which state transitions are independent from the duration of the previous state and the duration is only conditioned on the current state. Likewise, the CHSMM is different from a \textit{residential time HMM}~\cite{Yu10AI_HSMM}, where state transitions are assumed to be independent from the duration of the previous state. The independence assumption of these simplified models does not hold in practice as the duration of the current state is clearly dependent on the previous state as shown in Fig.~\ref{fig:dur_dist}b.

\subsection{Parameter Estimation}
As outlined in Section~\ref{sec:HSMM}, there are five parameters to be estimated: the number of states $N_S$, the maximum duration, the generalized state transition probabilities, the emission distribution, and the initial distribution.

As described in Section~\ref{sec:pre}, the number of states can be estimated by the empirical histogram of real power consumption --- the number of histogram peaks. Given the number of states, we apply K-means clustering to the real power trajectory and choose the longest duration as the maximum duration. The initial distribution can be easily estimated by the stationary distribution. In the following, we present with details on the estimation of transition and emission distributions.

To simplify the estimation of transition probabilities, we separate the transition from the generalized state  $(\tilde{x},\tilde{d})$ to $(\tilde{x}',\tilde{d}')$ conditioning on $\tilde{z}'$ into two parts: (1) the state transition and (2) the duration transition. The generalized state transition is given by
\begin{equation}
a_{(\tilde{x}, \tilde{d}),(\tilde{x}',\tilde{d}')}(\tilde{z}')=a^{S}_{\tilde{x}'}(\tilde{x},\tilde{d},\tilde{z}')a^{D}_{\tilde{d}'}(\tilde{x},\tilde{d},\tilde{x}',\tilde{z}')
\end{equation}
where 
\arraycolsep=1.4pt\def\arraystretch{2.2}
\begin{eqnarray}
a^{S}_{\tilde{x}'}(\tilde{x},\tilde{d},\tilde{z}')&=&\mathbb{P}[\tilde{x}'|\tilde{x}, \tilde{d},\tilde{z}']\\
a^{D}_{\tilde{d}'}(\tilde{x},\tilde{d},\tilde{x}',\tilde{z}')&=&\mathbb{P}[\tilde{d}'|\tilde{x}, \tilde{d},\tilde{x}',\tilde{z}'].\hspace{1.5em}
\end{eqnarray}
Note that this separation does not assume any independence of state or duration between two epoches. 

From the duration analysis in Section~\ref{sec:pre}, it is acknowledged that a single distribution is insufficient to describe the randomness and dependence of duration across different states. Another option is to estimate the transition distribution directly from data by using empirical frequency. This approach, however, suffers a major setback --- it cannot model a state transition which has never been observed in the past. To overcome these drawbacks, we propose a data-based approach which learns the transition probabilities via multinomial logistic regression (MNLR). The use of MRLR models arises from the fact that the state and duration are discrete variables. In particular, the state transition probability $a^{S}_{\tilde{x}'}(\tilde{x},\tilde{d},\tilde{z}')$ and the duration transition probability $a^{D}_{\tilde{d}'}(\tilde{x},\tilde{d},\tilde{x}',\tilde{z}')$ are defined in the following form:
\begin{equation}\label{eqn:mnlr_s}
a^{S}_{\tilde{x}'}(\tilde{x},\tilde{d},\tilde{z}')\triangleq\frac{e^{\alpha^1_{\tilde{x}'}\tilde{x}+\alpha^2_{\tilde{x}'}\tilde{d}+\alpha^3_{\tilde{x}'}\tilde{z}'}}{\sum_{\tilde{x}''\in\mathscr{S}} e^{\alpha^1_{\tilde{x}''}\tilde{x}+\alpha^2_{\tilde{x}''}\tilde{d}+\alpha^3_{\tilde{x}''}\tilde{z}'}}
\end{equation}
and 
\begin{equation}\label{eqn:mnlr_d}
a^{D}_{\tilde{d}'}(\tilde{x},\tilde{d},\tilde{x}',\tilde{z}')\triangleq\frac{e^{\beta^1_{\tilde{d}'}\tilde{x}+\beta^2_{\tilde{d}'}\tilde{d}+\beta^3_{\tilde{d}'}\tilde{x}'+\beta^4_{\tilde{d}'}\tilde{z}'}}{\sum_{\tilde{d}''\in\mathscr{D}} e^{\beta^1_{\tilde{d}''}\tilde{x}+\beta^2_{\tilde{d}''}\tilde{d}+\beta^3_{\tilde{d}''}\tilde{x}'+\beta^4_{\tilde{d}''}\tilde{z}'}}
\end{equation}
where $\alpha_{\tilde{x}}=(\alpha^1_{ {x}},\alpha^2_{ {x}},\alpha^3_{ {x}})$,  $\forall  {x}\in\mathscr{S}$, and $\beta_{d}=(\beta^1_{d},\beta^2_{d},\beta^3_{d},\beta^4_{d})$,  $\forall d\in\mathscr{D}$, are regression  coefficients. We note that the coefficients $\alpha^1_{ {x}},\alpha^2_{ {x}}, \beta^1_{d},\beta^2_{d},\beta^3_{d}$ are scalars, and $\alpha^3_{{x}}$ and $\beta^4_{d}$ are in compatible dimension with exogenous variable $z$. 

The emission probability $b(y|x,w)$ for the observed value $y$ at time $t$ conditioned on the current state $x$ and exogenous variable $w$ is assumed to be Gaussian:
\begin{equation}\label{dist:em}
y\sim\mathcal{N}\left(\gamma_x+ \phi w,\sigma^2\right)
\end{equation}
where $\gamma_x$ is the centroid associated with state $x$ from the K-means clustering algorithm. The values of the parameter $\alpha$, $\beta$, $\gamma$, $\phi$ and $\sigma$ are obtained using maximum likelihood estimation based on the states obtained from K-means clustering.

\subsection{State Oriented Model}\label{sec:state_MNLR}
The states of a CHSMM generally represent the operation of one or more appliance components (e.g. motors, heating elements). The duration distributions and transition probabilities associated with different states may depend on different physical processes and be influenced by human behavior in different ways. Therefore, they may vary significantly from one state to another. Model bias may be high if the duration distributions of all states and all transition probabilities are each estimated by a single MNLR model. As a variation of the proposed CHSMM, we estimate the duration distributions of each state and the transition probabilities from each state to all other states using different MNLR models. 

\subsection{Weighted Logistic Regression}\label{sec:weighted_MNLR}
The second model variation applies to appliances with a heavy-tailed duration distribution. For an appliance such as an air conditioner (A/C), the probabilities associated with unusually long duration events are low and the relative number of such events is small. For example, the compressor in an A/C may typically have an ON duration of approximately ten minutes, but occasionally run for several hours during peak thermal load conditions. These rare events can be of particular interest to utilities, for example for peak load management, but difficult to learn using basic MNLR due to the class imbalance. To address this, we perform weighted MNLR, where the data for each generalized state is weighted as a function of the duration $d_k$ and an integer-valued weighting factor $a$
\begin{equation}
m_{k}=1+\frac{\tilde{d}_{k}}{a}
\end{equation}

\section{Short-Term Load Forecasting}

In this section, we describe an algorithm for short-term load forecasting using the proposed CHSMM. Suppose at time $t$, we want to predict the power trajectory from $t+1$ to $t+H$, where $H$ is the prediction horizon. 

The first step is to predict how long the appliance remains in the current state. Let $k$ be the index of the epoch at time $t$, such that  $\tilde{x}_k$ is the state and $\tilde{d}_k$ is the duration. Given the observation of the previous generalized state $(\tilde{x}_{k-1},\tilde{d}_{k-1})$ and the exogenous variable $\tilde{z}_k$, the prediction $\hat{\tilde{d}}_k$ of the duration $\tilde{d}_k$ at epoch $k$ is given by
\begin{equation}\label{eqn:initial_d}
\hat{\tilde{d}}_k=\underset{{d}\in\mathscr{D},{d}\geq t-t^s_{k}+1}{\arg\!\max} \left\{ a^{D}_{d}\left(\tilde{x}_{k-1},\tilde{d}_{k-1},\tilde{x}_k,\tilde{z}_{k}\right)\right\}
\end{equation}
where $t^e_{k-1}$ is the final timestep of state $\tilde{x}_{k-1}$.

The constraint ${d}\geq t-t^s_{k}+1$ restricts the duration $\tilde{d}_k$ of the current state $\tilde{x}_k$ to be at least $t-t^s_{k}+1$, since the last state transition happened at time $t^s_{k}=t^e_{k-1}+1$.

Given $\hat{\tilde{d}}_k$, we can predict the most likely state trajectory until time $t+H$ via an iterative procedure using the state and duration transition probabilities given in (\ref{eqn:mnlr_s}-\ref{eqn:mnlr_d}). In particular, starting from time $t^e_{k}+1$, we maximize the state transition probability to find the most likely state $\hat{\tilde{x}}_{k+1}$. Using this prediction, we find the most likely duration $\hat{\tilde{d}}_{k+1}$, and repeat this process until time $t+H$.

Finally, the observation trajectory $(y_{t+1},y_{t+2},\cdots, y_{t+H})$ can be obtained from the emission distribution given in (\ref{dist:em}) using the predicted state trajectory and associated exgonous variables. 

This prediction procedure is summarized in Algorithm~\ref{alg:forecast}.
\begin{algorithm}
\caption{Short-term forecast using CHSMM.}\label{alg:forecast}
\begin{algorithmic}[1]
\REQUIRE previous generalized state $(\tilde{x}_{k-1},\tilde{d}_{k-1})$, current state $\tilde{x}_k$,  probability distribution $a^S(\cdot)$, $a^D(\cdot)$, and $b(\cdot)$, and predicted trajectory of exogenous variable $\hat{z}_{t+1}, \cdots, \hat{z}_{t+H}$ and $\hat{w}_{t+1}, \cdots, \hat{w}_{t+H}$.
\STATE  \textbf{initialize} $n \gets k$, $\tau \gets t$, $\hat{\tilde{d}}_n=\hat{\tilde{d}}_k$ given in (\ref{eqn:initial_d}), 
\WHILE {$\tau \leq t+H$}
\STATE $n \gets n+1$
\STATE $\hat{\tilde{x}}_n=\underset{x\in\mathscr{S}}{\arg\!\max} \left\{ a^{S}_{x}\left(\hat{\tilde{x}}_{n-1},\hat{\tilde{d}}_{n-1},\hat{\tilde{z}}_{n}\right)\right\}$
\STATE $\hat{\tilde{d}}_n=\underset{d\in\mathscr{D}}{\arg\!\max} \left\{ a^{D}_{d}\left(\hat{\tilde{x}}_{n-1},\hat{\tilde{d}}_{n-1},\hat{\tilde{x}}_n,\hat{\tilde{z}}_{n}\right)\right\}$
\STATE $\tau \gets \max{\left(\sum_{i=k}^n \hat{\tilde{d}}_{i}, t+H \right)}$
\STATE $\hat{y}_{s}=\mathbb{E}[y_s|\hat{\tilde{x}}_n, \hat{w}_s]$,  $\forall s\in \left[\tau- \hat{\tilde{d}}_{n}+1, \tau\right]$
\ENDWHILE
\STATE \textbf{result} predicted trajectory $\hat{y}_{t+1}, \cdots,\hat{y}_{t+H}$
\end{algorithmic}
\end{algorithm}

\section{Case Study Methodology}\label{sec:case_study}
We validated the performance of the CHSMM model for load prediction using real-world appliance power data. In this section, we describe the data and model parameters for each appliance type and define metrics for evaluating load forecasting performance.

\subsection{Data}
Appliance-level real power data with 1-minute sampling resolution was obtained from the Pecan Street database~\cite{PecanSt17Dataport}. We analyzed four different types of residential loads: A/Cs, refrigerators, pool pumps, and EV charging. In particular, we performed load prediction for 50 A/Cs, 20 refrigerators, 20 pool pumps, and 20 EVs. We used data from 2016 to train the model and data from 2015 to test the load prediction. For A/Cs, pool pumps, and refrigerators, data over the same ten week period was used for training and testing (June 23 - Sep 1) to eliminate any seasonal effects on load behavior. Since the relative frequency of EV charging is lower than that of the other appliances, a longer training period (Apr 14 - Oct 4) was used. Hourly outdoor temperature for Austin, Texas was also obtained from the Pecan Street database~\cite{PecanSt17Dataport}.
\subsection{Parameter Specification}
The number of clusters in the K-means algorithm for state abstraction was manually chosen based on the number of peaks in the histogram of real power from the training data. The heuristic elbow method~\cite{tibshirani2001estimating} was also implemented, but it usually yielded fewer clusters, due to the low frequency of certain states. 

For A/Cs, the state and duration transition probabilities were conditioned on the outdoor temperature and time of day. For the other appliances, the state and duration transition probabilities were only conditioned on the time of day. For the emission distribution, the power consumption of A/Cs in the ON state was approximated using linear regression, using the outdoor temperature as a single feature. 

For the other appliances, the power consumption in each state is relatively constant over time. Therefore the emission distribution was not conditioned on exogenous variables. Parameters of the MNLR models were estimated via maximum likelihood estimation, which was implemented using the Scikit-Learn machine learning package in Python ~\cite{scikit-learn}.

\subsection{Performance metric}
We used the normalized root mean square error (NRMSE) as a performance metric. The NRMSE of an aggregation of $N$ loads over prediction horizon $[t+1, t+H]$ is defined as 
\begin{equation}\small
    \text{NRMSE}=\frac{\sqrt{\frac{1}{H}\left(\sum_{i=1}^N\sum_{\tau=t+1}^{t+H} \left(y^{(i)}_\tau - \hat{y}^{(i)}_\tau \right)\right)^2}}{\max\left(\sum_{i=1}^N\sum_{\tau=t+1}^{t+H} y^{(i)}_\tau\right)-\min\left(\sum_{i=1}^N\sum_{\tau=t+1}^{t+H} y^{(i)}_\tau\right)}
\end{equation}
where $y_\tau^{(i)}$ and $\hat{y}_\tau^{(i)}$ are the actual and predicted power of the load $i$ at time $\tau$ respectively.

\section{Case Study Results}\label{sec:case_study_results}

In this section, we present case study results from the load prediction of individual appliances and aggregations of appliances of the same type. Prediction performance is compared with that of the benchmark technique --- the basic HSMM described in Section~\ref{sec:HSMM}. Weighted MNLR and state-specific modeling for A/Cs are also implemented, showing an improvement in prediction performance. Finally, we demonstrate the use of the model for anomaly detection.

\subsection{Load Forecasting for Individual Appliances}\label{sec:short_forecast_ind}
\begin{figure}
\includegraphics[width=.48\textwidth]{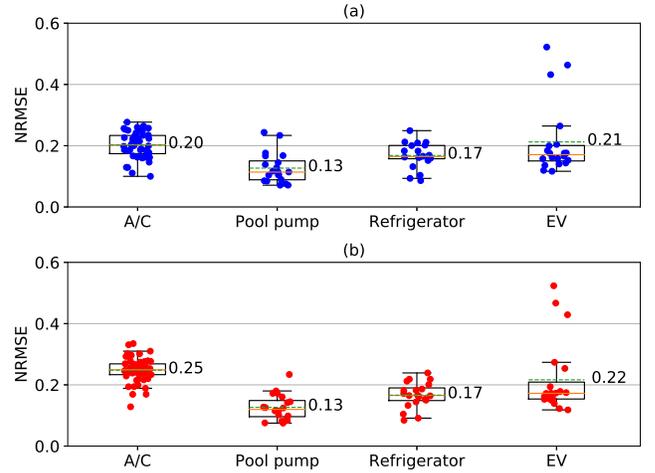}
\caption{NRMSE of 1-hour ahead load prediction for individual appliances using (a) CHSMM and (b) HSMM.}\label{fig:boxplot_NRMSE}
\end{figure}

Performance of the load prediction algorithm tends to vary for appliances of the same type due to heterogeneity in the sample population. This can be observed in Fig.~\ref{fig:boxplot_NRMSE}, which shows individual 1-hour ahead load prediction errors for different appliance types for both the CHSMM and HSMM. This heterogeneity arises from differences between appliances of the same type and variation in consumer behavior between households.

\begin{figure}
\includegraphics[width=.48\textwidth]{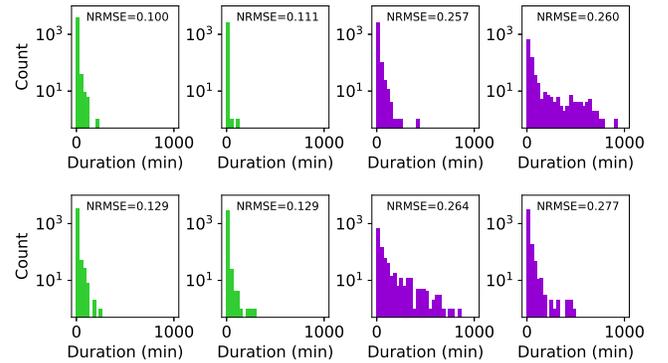}
\caption{Histogram of the duration of the ON state from training data for A/Cs with the lowest (green) and highest (purple) NRMSE values. }\label{fig:AC_duration}
\end{figure}

The models for A/Cs have two states: compressor ON and compressor OFF. For both the CHSMM and the CHMM, the load prediction accuracy tends to be worse for appliances with heavy-tailed duration distributions. This can be observed in Fig.~\ref{fig:AC_duration}, which shows histograms of the duration of the ON state for A/Cs with four lowest NRMSE values (green) and the four highest NRMSE values (purple). In general, the duration distribution of each state depends on the cycling frequency of the compressor, which depends on the thermal parameters of the home and the sizing of the compressor relative to the time-varying thermal load. A/Cs with heavy-tailed duration distributions may be undersized or experience frequent changes in the thermostat setpoint. As shown in Fig.~\ref{fig:boxplot_NRMSE}, the mean NRMSE for the CHSMM is lower than that of the HSMM by approximately 0.05, showing that there is a significant improvement in load prediction from incorporating exogenous variables into the model. Since the power consumption of the compressor ON state tends to vary by more than 0.5 kW as a linear function of the outdoor temperature, conditioning the emission distribution on the outdoor temperature improves the model substantially.
			
The models for pool pumps have 2-9 operating states. Models with 2 states represent single speed pumps and models with 3+ states represent variable speed pumps, where each state likely corresponds with a different pump speed and/or the operation of an additional secondary load. Residential pool pumps generally have a timer and preset settings for different daily schedules. Therefore, most of the power profiles of the pool pumps in the training data were highly consistent from day to day. The mean NRMSE for pool pumps was lower than the mean NRMSE of all other load types for both the CHSMM and the HSMM, likely due to this consistency. The difference in the mean NRMSE between the CHSMM and the HSMM was negligible, indicating that including the hour of the day in the model does not improve performance. This suggests that the conditional probabilities describing the generalized state transition structure of the HSMM are sufficient to model the diurnal patterns of pool pump operation. Additionally, the maximum load prediction horizon considered in this study was 1 hour. Had longer prediction horizons been considered, it is likely that the CHSMM would result in larger improvements in performance. 

All models for refrigerators have either 3 or 4 states. In general, these states represent the two operating modes of the compressor (ON and OFF), the defrost cycle, and potentially an icemaker cycle, depending on the refrigerator type. Less frequent state transitions, such as to the defrost cycle, tend to be predicted with lower accuracy. Similar to the results for pool pumps, there was a negligible difference in the mean NRMSE between the HSMM and the CHSMM. This shows that conditioning the generalized state transition probabilities on the hour of the day does not improve load prediction performance. In general, the cycling rate of the refrigerator compressor and the frequency of the defrost cycle depends on the frequency of refrigerator door openings, which affect the temperature and relative humidity in the conditioned space. Incorporating the current temperature and relative humidity into the model as exogenous variables would likely improve the prediction accuracy of the CHSMM. 

Clustering analysis indicated that most EVs in the sample population were equipped with either 3.3 or 6.6 kW chargers. In all cases, the magnitude of the charging power was relatively constant over time such that the load could be well approximated with a 2-state model. There was a small improvement from incorporating the hour of the day into the model (reduction of the mean NRMSE of 0.0042). As EV charging patterns are generally time-dependent, it is likely that the CHSMM would result in larger improvements over the HSMM for longer prediction horizons. 

\begin{figure}
\includegraphics[width=.48\textwidth]{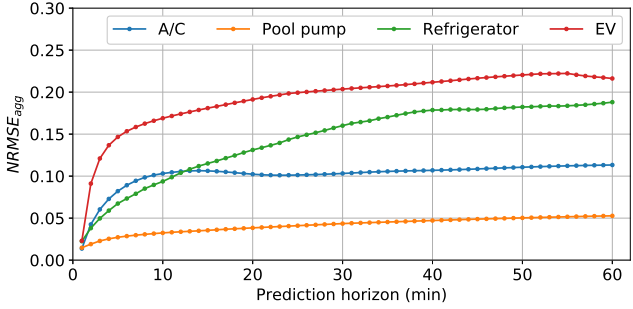}
\caption{NRMSE for aggregations of 20 appliances modeled with the CHSMM as a function of the prediction horizon. }\label{fig:agg_NRMSE}
\end{figure}

\subsection{Load Forecasting for Aggregated Loads}\label{sec:short_forecast_agg}

For certain applications, it may be necessary to predict the total load of an aggregation of appliances, rather an individual appliance. The NRMSE calculated for aggregations of appliances of the same type are shown in Fig.~\ref{fig:agg_NRMSE}. There are at least two factors which affect the change in the prediction error with respect to the prediction horizon. First, uncertainty in the predicted power trajectory becomes larger as the prediction horizon increases, which generally results in an increase in the NRMSE. Second, since the NRMSE is calculated using the sum of the discrete power trajectory over the prediction horizon, increasing the prediction horizon can reduce prediction error since positive and negative deviations between the actual and predicted power trajectories tend to average out. For all loads besides A/Cs, the first factor is dominant, resulting in an increase in the NRMSE with respect to the prediction horizon. 

\begin{figure}
\includegraphics[width=.48\textwidth]{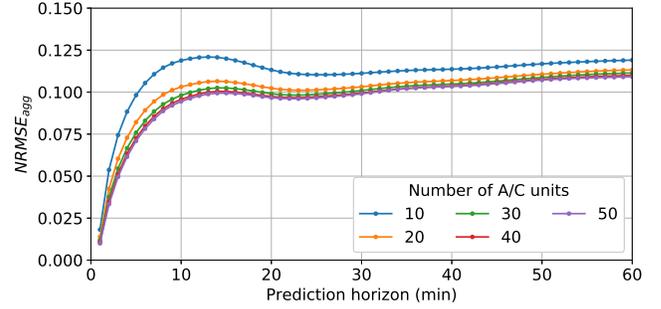}
\caption{The NRMSE for different levels of aggregation for A/Cs modeled with the CHSMM for a range of prediction horizons. }\label{fig:AC_aggregation}
\end{figure}

\begin{figure}
\includegraphics[width=.48\textwidth]{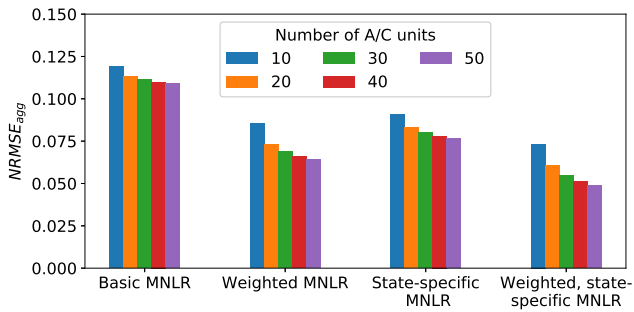}
\caption{The aggregated NRMSE for load prediction of A/Cs using different refinements of the CHSMM: (a) basic model, (b) weighted MNLR, (c) state-specific MNLR, and (d) weighted, state-specific MNLR. All results are evaluated with a 60 minute prediction horizon.}\label{fig:model_variations}
\end{figure}

We evaluated the effect of load aggregation on load prediction error for A/Cs by varying the number of loads in the aggregation. As shown in Fig.~\ref{fig:model_variations}, there is a reduction in error of 0.01-0.02 when the size of the aggregation is increased from 10 to 20 loads. Larger aggregations of loads result in a negligible reduction in prediction error.

\subsection{Model Refinements for Air Conditioners}\label {sec:short_forecast_refine}
In this section, we present results of the two variations of the proposed CHSMM for A/Cs: weighted MNLR and state-specific MNLR. Fig.~\ref{fig:AC_aggregation} shows the load prediction error for different levels of aggregation for four cases: (a) the basic model, (b) weighted MNLR, (c) state-specific MNLR, and (d) both refinements implemented together. While each refinement reduces the error, implementing both methods together results in the most significant improvement. For this case, the NRMSE is reduced from 0.109 to 0.050 for 1-hour ahead prediction with aggregation of 50 A/Cs. Weighted MNLR improves the prediction accuracy of infrequent long duration events which occur when the compressor cycles for extended periods of time during high thermal loads. The increase in model complexity from using state-specific models reduces the prediction error by achieving a better approximation of the duration distributions of the ON and OFF states.

\subsection{Prediction-Based Anomaly Detection}
Load prediction is based on the assumption that observed patterns reoccur in the future. If this assumption does not hold true, the predicted values may be far off from measured values. Additionally, observations that are vastly different from expected values tell us that the model used does not explain the observed values. In theory, there may be two reasons for this. First, the structure of the model may not be appropriate for modeling the behavior of the system, and second, the behavior of the system may actually be deviating from the expected and explainable behavior. Since we observed low mean load prediction errors for the vast majority of appliances modeled with the CSHMM, we can assume that the fitted models describe typical load behavior fairly well. Therefore, unusually high load prediction error can be used as an indication of anomalous behavior, either with respect to the typical behavior of that appliance or other appliances of the same type.

\begin{figure}
\includegraphics[width=.48\textwidth]{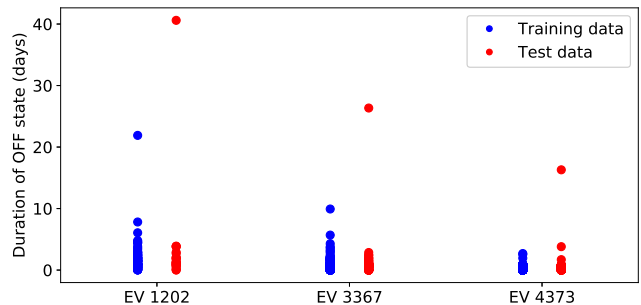}
\caption{Durations of the OFF state for the three EVs with the highest prediction error, for both the training and testing data.}\label{fig:duration_EV}
\end{figure}

Three examples of EVs with high load prediction error can be observed in Fig.~\ref{fig:boxplot_NRMSE}. Prediction error was large in these cases because there was a lapse in EV charging that lasted more than two weeks. In particular, the model became inaccurate when the longest duration associated with the OFF state was much larger for the test data than for the training data. This can be observed in Fig.~\ref{fig:duration_EV}, which shows the range of durations of the OFF state for the training and test data for the three EVs with the largest error. For EV 4373, which had the largest error, the longest duration for the OFF state was 0.98 days for the training data and 16.3 days for the testing data. This could occur if homeowners generally charge their vehicle on a daily basis but occasionally travel away from home for an extended period of time. These periods cannot be easily predicted by the algorithm, given the exogenous variables that are included in the model.

\section{Conclusion}\label{sec:con}

In this paper, we propose a data-driven approach for high resolution load modeling and short-term forecasting of residential appliances, based on a CHSMM.  The probabilistic model is based on two primary assumptions: that appliances are characterized by a finite set of operating states, and the duration in each state is a random variable explicitly defined by a potentially complex duration distribution. In addition, to account for behavioral and environmental factors, the state transition and duration probability distributions are conditioned on exogenous variables. We utilize scalable statistical learning methods to learn model parameters directly from historical sub-metered power consumption data, allowing for efficient implementation. 

Additionally, we developed a short-term load forecasting algorithm using the learned load model. The model and load prediction algorithm are generally applicable to most residential appliances, however results show that small adjustments in the parameter estimation methodology can improve load prediction performance for specific appliance types, such as A/Cs. Case studies performed using 1-minute resolution data from real homes show the effectiveness of the proposed model and forecasting algorithm for different levels of aggregation. Finally, as an additional application, we present an example of how the methodology can be used for anomaly detection.

\bibliographystyle{IEEEtran}
{\bibliography{ref}}
\end{document}